\pdfoutput=1

\documentclass[12pt]{article}

\usepackage{amsfonts}
\usepackage{amsmath}
\usepackage{amssymb}
\usepackage{mathrsfs}
\usepackage{bigints}
\usepackage{booktabs}
\usepackage[nosort]{cite}
\usepackage{color}
\usepackage{dsfont}
\usepackage{float}
\usepackage{framed}
\usepackage{graphicx}
\usepackage{indentfirst}
\usepackage{mathrsfs}
\usepackage{multirow}
\usepackage{pdflscape}
\usepackage{setspace}
\usepackage{subdepth}
\usepackage{subfig}
\usepackage{titlesec}
\usepackage[dotinlabels]{titletoc}
\usepackage{wrapfig}
\usepackage[all]{xy}
\usepackage{young}
\usepackage[vcentermath]{youngtab}
\usepackage{relsize}
\usepackage{stackengine}
\usepackage{datetime}

\usepackage{hyperref}

\numberwithin{equation}{section}

\usepackage{verbatim}

\newcommand{\be}{\begin{equation}}
\newcommand{\ee}{\end{equation}}

\newcommand{\bml}{\begin{multline}}
\newcommand{\emll}{\end{multline}}

\newcommand{\nn}{\nonumber}

\def\({\left(} \def\){\right)}
\def\[{\left[} \def\]{\right]}

\def\mA{\mathcal{A}}

\def\v{\vec}

\def\g{\gamma}

\def\lam{\lambda}

\def\o{\omega}

\newcommand{\bea}{\begin{eqnarray}}
\newcommand{\eea}{\end{eqnarray}}

\newcommand{\order}[1]{\mathcal{O}\!\left(#1\right)}

\usepackage[left=2cm,right=2cm,top=2cm,bottom=2cm]{geometry}
\titleformat{\section}{\large\bfseries}{\thesection.}{4pt}{}
\titlespacing{\section}{0pt}{22pt}{6pt}

\titleformat{\subsection}{\normalfont\bfseries}{\thesubsection.}{4pt}{}
\titlespacing{\subsection}{0pt}{18pt}{6pt}

\titleformat{\subsubsection}{\normalfont\itshape}{\thesubsubsection.}{4pt}{}
\titlespacing{\subsubsection}{0pt}{16pt}{6pt}

\def\ie{\begin{equation}\begin{aligned}}
\def\fe{\end{aligned}\end{equation}}

\def\tilde{\widetilde}

\def\1{{\mathds 1}}

\def\mA{\mathcal{A}}

\def\mL{\mathcal{L}}

\def\bf{\mathbf}

\def\o{\omega}
\def\v{\vec }

\DeclareFontShape{OT1}{cmr}{mx}{n}%
    {<->cmr10}{}
\newcommand{\mytitlefont}{\fontseries{mx}\selectfont}
\DeclareMathAlphabet{\titlemath}{OT1}{cmr}{mx}{n}

\newcommand{\bi}{\begin{itemize}}
\newcommand{\ei}{\end{itemize}}

\usepackage{bm}

\usepackage{tikz}
\tikzset{every picture/.style={line width=0.75pt}} 

\usetikzlibrary{decorations.markings,arrows.meta}

\tikzset{
  midarrow/.style={
    postaction={
      decorate,
      decoration={
        markings,
        mark=at position 0.5 with {\arrow{Stealth[length=8pt,width=8pt]}}
      }
    }
  }
}

\newcommand{\HSdiagram}[1]{%
  \resizebox{#1}{!}{%
    \begin{tikzpicture}[line width=0.8pt]
      \coordinate (v1) at (0,0);
      \coordinate (v2) at (4,0);
      \draw[midarrow] (-1.7,1.3) -- (v1);
      \draw[midarrow] (-1.7,-1.3) -- (v1);
      \draw[midarrow] (v2) -- (5.7,1.3);
      \draw[midarrow] (v2) -- (5.7,-1.3);
      \node at (-1.85,1.55) {$2$};
      \node at (-1.85,-1.55) {$1$};
      \node at (5.85,1.55) {$4$};
      \node at (5.85,-1.55) {$3$};
      \draw[dashed] (v1) -- (1.75,0);
      \draw[dashed] (2.25,0) -- (v2);
      \fill[gray!50] (2,0) circle (0.27);
      \draw (2,0) circle (0.27);
    \end{tikzpicture}%
  }%
}

\newcommand{\HSSdiagram}[1]{%
  \resizebox{#1}{!}{%
    \begin{tikzpicture}[line width=0.8pt]
      \coordinate (v1) at (0,0);
      \coordinate (v2) at (4,0);
      \draw[midarrow] (v1) -- (-1.7,1.3);
      \draw[midarrow] (-1.7,-1.3) -- (v1);
      \draw[midarrow] (v2) -- (5.7,1.3);
      \draw[midarrow] (5.7,-1.3) -- (v2);
      \node at (-1.85,1.55) {$3$};
      \node at (-1.85,-1.55) {$1$};
      \node at (5.85,1.55) {$4$};
      \node at (5.85,-1.55) {$2$};
      \draw[dashed] (v1) -- (1.75,0);
      \draw[dashed] (2.25,0) -- (v2);
      \fill[gray!50] (2,0) circle (0.27);
      \draw (2,0) circle (0.27);
    \end{tikzpicture}%
  }%
}

\newcommand{\Pidiagram}[1]{%
  \resizebox{#1}{!}{%
    \begin{tikzpicture}[line width=0.8pt,baseline={(current bounding box.center)}]
      \coordinate (L) at (0,0);
      \coordinate (R) at (3,0);

      \node[font=\Large] at (-0.75,0) {$\Pi =$};

      \draw[dashed] (L) -- (0.4,0);
      \draw[dashed] (2.6,0) -- (R);

      \draw[midarrow]
        (0.4,0)
        .. controls (0.4,0.9) and (2.6,0.9) .. (2.6,0);

      \draw[midarrow]
        (2.6,0)
        .. controls (2.6,-0.9) and (0.4,-0.9) .. (0.4,0);
    \end{tikzpicture}%
  }%
}

\newcommand{\Sigmadiagram}[1]{%
  \resizebox{#1}{!}{%
    \begin{tikzpicture}[line width=0.8pt,baseline={(0,0)}]
      \coordinate (L) at (0.6,0);
      \coordinate (R) at (3.4,0);
      \coordinate (a) at (1,0);
      \coordinate (b) at (3,0);

      \node[font=\Large] at (-0.05,0) {$\Sigma =$};
      \draw[midarrow] (L) -- (R);

      \draw[dashed] (a)
        .. controls (1.15,1.15) and (2.85,1.15) .. (b);

      \fill[gray!50] (2,0.8625) circle (0.27);
      \draw (2,0.8625) circle (0.27);
    \end{tikzpicture}%
  }%
}

\newcommand{\BubbleChainTwo}[1]{%
  \resizebox{#1}{!}{%
    \begin{tikzpicture}[line width=0.8pt,baseline={(current bounding box.center)}]
      \coordinate (v1) at (0,0);
      \coordinate (v2) at (5.7,0);

      \draw[midarrow] (-1.5,1.3) -- (v1);
      \draw[midarrow] (-1.5,-1.3) -- (v1);
      \draw[midarrow] (v2) -- (7.2,1.3);
      \draw[midarrow] (v2) -- (7.2,-1.3);

      \node at (-1.65,1.55) {$2$};
      \node at (-1.65,-1.55) {$1$};
      \node at (7.35,1.55) {$4$};
      \node at (7.35,-1.55) {$3$};

      \draw[midarrow]
        (v1) .. controls (0.15,0.75) and (1.35,0.75) .. (1.5,0);
      \draw[midarrow]
        (v1) .. controls (0.15,-0.75) and (1.35,-0.75) .. (1.5,0);

      \draw[midarrow]
        (1.5,0) .. controls (1.65,0.75) and (2.85,0.75) .. (3,0);
      \draw[midarrow]
        (1.5,0) .. controls (1.65,-0.75) and (2.85,-0.75) .. (3,0);

      \node[font=\Large] at (3.75,0) {$\cdots$};

      \draw[midarrow]
        (4.2,0) .. controls (4.35,0.75) and (5.55,0.75) .. (5.7,0);
      \draw[midarrow]
        (4.2,0) .. controls (4.35,-0.75) and (5.55,-0.75) .. (5.7,0);

    \end{tikzpicture}%
  }%
}

\newcommand{\BubbleChainOne}[1]{%
 \resizebox{#1}{!}{%
    \begin{tikzpicture}[line width=0.8pt,baseline={(current bounding box.center)}]
      \coordinate (v1) at (0,0);
      \coordinate (v2) at (5.7,0);

      \draw[midarrow] (v1) -- (-1.5,1.3);       
      \draw[midarrow] (-1.5,-1.3) -- (v1);      
      \draw[midarrow] (v2) -- (7.2,1.3);        
      \draw[midarrow] (7.2,-1.3) -- (v2);       

      \node at (-1.65,1.55) {$3$};
      \node at (-1.65,-1.55) {$1$};
      \node at (7.35,1.55) {$4$};
      \node at (7.35,-1.55) {$2$};

      \draw[midarrow]
        (v1) .. controls (0.15,0.75) and (1.35,0.75) .. (1.5,0);
      \draw[midarrow]
        (1.5,0) .. controls (1.35,-0.75) and (0.15,-0.75) .. (v1);

      \draw[midarrow]
        (1.5,0) .. controls (1.65,0.75) and (2.85,0.75) .. (3,0);
      \draw[midarrow]
        (3,0) .. controls (2.85,-0.75) and (1.65,-0.75) .. (1.5,0);

      \node[font=\Large] at (3.75,0) {$\cdots$};

      \draw[midarrow]
        (4.2,0) .. controls (4.35,0.75) and (5.55,0.75) .. (5.7,0);
      \draw[midarrow]
        (5.7,0) .. controls (5.55,-0.75) and (4.35,-0.75) .. (4.2,0);

    \end{tikzpicture}%
  }%
}

\newcommand{\SigmadiagramM}[1]{%
  \resizebox{#1}{!}{%
    \begin{tikzpicture}[line width=0.8pt,baseline={(current bounding box.center)}]

      \coordinate (v1) at (0,0);
      \coordinate (v2) at (1.8,0);

      \draw (-0.3,0) -- (v1);
      \draw (v2) -- (2.1,0);

      \draw[midarrow] (v1) -- (v2);

      \draw[midarrow]
        (v1) .. controls (0.2,0.85) and (1.6,0.85) .. (v2);

      \draw[midarrow]
        (v2) .. controls (1.6,-0.85) and (0.2,-0.85) .. (v1);

      \node at (-0.45,0.35) {$1$};
      \node at (2.25,0.35) {$2$};

    \end{tikzpicture}%
  }%
}

\newcommand{\SigmadiagramMM}[1]{%
\resizebox{#1}{!}{%
    \begin{tikzpicture}[line width=0.8pt,baseline={(current bounding box.center)}]

      \coordinate (v1) at (0,0);
      \coordinate (v2) at (1.8,0);

      \draw[dashed] (-0.6,0) -- (v1);
      \draw[dashed] (v2) -- (2.4,0);

      \draw[midarrow] (v1) -- (v2);

      \draw[midarrow]
        (v1) .. controls (0.2,0.85) and (1.6,0.85) .. (v2);

      \draw[midarrow]
        (v2) .. controls (1.6,-0.85) and (0.2,-0.85) .. (v1);

      \node at (-0.75,0.35) {$1$};
      \node at (2.55,0.35) {$2$};

    \end{tikzpicture}%
  }%
}

\newcommand{\SigmadiagramMMM}[1]{%
\resizebox{#1}{!}{%
    \begin{tikzpicture}[line width=0.8pt,baseline={(current bounding box.center)}]

      \coordinate (v1) at (0,0);
      \coordinate (v2) at (1.8,0);
      \coordinate (vm) at (0.9,0);

      \draw[dashed] (-0.6,0) -- (v1);
      \draw (v2) -- (2.4,0);

      \draw (v1) -- (vm);
      \draw[dashed] (vm) -- (v2);
      \path[midarrow] (v1) -- (v2);

      \draw[midarrow]
        (v1) .. controls (0.2,0.85) and (1.6,0.85) .. (v2);

      \draw[midarrow]
        (v2) .. controls (1.6,-0.85) and (0.2,-0.85) .. (v1);

      \node at (-0.75,0.35) {$1$};
      \node at (2.55,0.35) {$2$};

    \end{tikzpicture}%
  }%
}

\newcommand{\SigmadiagramMMMM}[1]{%
\resizebox{#1}{!}{%
    \begin{tikzpicture}[line width=0.8pt,baseline={(current bounding box.center)}]

      \coordinate (v1) at (0,0);
      \coordinate (v2) at (1.8,0);

      \draw[dashed] (-0.6,0) -- (v1);
      \draw (v2) -- (2.4,0);

      \draw[midarrow] (v1) -- (v2);

      \draw[midarrow]
        (v1) .. controls (0.2,0.85) and (1.6,0.85) .. (v2);

      \draw (0.9,-0.6375) .. controls (0.55,-0.6375) and (0.2,-0.425) .. (v1);
      \draw[dashed] (v2) .. controls (1.6,-0.425) and (1.25,-0.6375) .. (0.9,-0.6375);
      \path[midarrow]
        (v2) .. controls (1.6,-0.85) and (0.2,-0.85) .. (v1);

      \node at (-0.75,0.35) {$1$};
      \node at (2.55,0.35) {$2$};

    \end{tikzpicture}%
  }%
}
\begin{document}

\begin{titlepage}

\begin{center}

~\\[1cm]

{\fontsize{20pt}{0pt} \mytitlefont Kinetic Equations and Turbulence in Large $N$ Vector and SYK-like Models }\\[10pt]

~\\[0.2cm]

{\fontsize{14pt}{0pt}Vladimir Rosenhaus and Daniel Schubring}

~\\[0.1cm]

\it{ Initiative for the Theoretical Sciences}\\ \it{ The Graduate Center, CUNY}\\ \it{
 365 Fifth Ave, New York, NY 10016, USA}\\[.5cm]

~\\[0.6cm]

\end{center}

\noindent 

We apply large $N$ expansion techniques to derive kinetic equations for strongly coupled, far-from-equilibrium field theories with $O(N)$ symmetry. For large $N$ vector models, the quasi-classical approximation holds. Incorporating arbitrary momentum-dependent interactions, we show the dynamics obey the standard kinetic equation, with the scattering amplitude given by the retarded propagator of the auxiliary Hubbard-Stratonovich field. This has broad applications to systems like spin wave and Langmuir ion-acoustic turbulence. For large $N$ random coupling (SYK-like) models in $d$ dimensions, leading-order melon diagrams induce order-one propagator renormalization and finite spectral width. The corresponding kinetic equation, unlike those for standard weakly interacting systems, does not have a local-in-time collision integral. We find evidence for a stationary, scale-invariant infrared solution, revealing a potentially novel strong turbulence scaling exponent. 

\vfill 

\today
\end{titlepage}

\tableofcontents
~\\[-20pt]

\vspace{.3cm}

\section{Introduction}

The large $N$ expansion, in which one generalizes the theory to have a large number $N$ of fields and solves  perturbatively in $1/N$ and to all orders in the coupling, is a powerful technique for studying quantum field theories at strong coupling. 

Our goal is to apply large $N$ techniques to the study of strongly coupled field theories that are far from equilibrium. We will study two classes of models which contain $O(N)$ symmetry: vector models and random coupling (SYK-like) models. For each, we derive the kinetic equation, which is the appropriate description of a statistical ensemble of many interacting particles or waves. 

In the vector model case, discussed in Sec.~\ref{Sec2}, we consider arbitrary momentum-dependent couplings, finding a simple and elegant result that generalizes the case of constant coupling \cite{Kraichnan, Nowak:2011sk,B15,Gaz19,bergesGasenzerScheppach2010, berges2002,bergesRothkopfSchmidt2008, Rosenhaus:2024iqw, RF2, Noel:2026clb} and has applications to a wide range of systems, such as spin wave turbulence and Langmuir ion-acoustic wave turbulence \cite{ZLF}. In the random coupling case, discussed in Sec.~\ref{Sec3}, we derive a qualitatively new type of kinetic equation, which has stationary scale-invariant (far-from-equilibrium) solutions with a new scaling exponent for strong turbulence. 

We begin by collecting the essential ingredients of the Keldysh formalism. 
The Schwinger-Dyson equation defines the propagator in terms of the free propagator and the self-energy, 
\be \label{11}
G^{-1}(\o, \bf k) = G_0^{-1}(\o, \bf k) - \Sigma(\o,\bf k)~.
\ee
Quantum field theory naturally computes the in-out correlation functions that are appropriate for the $S$-matrix. To get an expectation value (an in-in correlator) one evolves forwards in time, acts with the operator, and then evolves backwards in time to the initial time. It is convenient to split the field into two fields, one on each branch of this Keldysh contour. As a result, the propagators become two by two matrices, see e.g. \cite{Hu:2025bqi},
\be \label{12}
G= \begin{pmatrix} G^K & G^R \\ G^A & 0\end{pmatrix}~, \quad~  \quad \Sigma = \begin{pmatrix} 0 & \Sigma^A \\ \Sigma^R & \Sigma^K \end{pmatrix}~,
\ee
where $G^R$ and $G^K$ are defined as
\be \label{13}
G^R(1, 2) = \theta(t_1 - t_2) \langle [ \Psi(1), \Psi^\dagger(2) ] \rangle, \quad G^K(1, 2) =  \langle \{ \Psi(1), \Psi^\dagger(2) \} \rangle~,
\ee
where our scalar field is $\Psi$,   $1$ denotes $(t_1, x_1)$ and $2$ denotes $(t_2, x_2)$. Note that $G^A(1, 2) = G^R(2, 1)^*$.
Inserting  (\ref{11}) into (\ref{12}) gives, 
\be \label{14}
G^R(\omega, \bf k)^{-1} = G^R_0(\omega, \bf k)^{-1}-\Sigma^R(\o,\bf k)~, \ \ \ \ G^K(\omega, \mathbf{k}) = G^R(\omega, \mathbf{k}) \Sigma^K(\omega, \mathbf{k}) G^A(\omega, \mathbf{k})~.
\ee

\section{Vector models} \label{Sec2}
We will be studying large $N$ vector models. For concreteness, we take a nonrelativistic complex field $\Psi$. There are two natural vector generalizations of the nonlinear Schr\"odinger equation: 
\be \label{case1}
\mL =  i \vec{\Psi}^* {\cdot} \partial_t \vec{\Psi} - |\nabla \vec{\Psi}|^2 - \frac{\lambda}{2N} (\vec{\Psi}^* \cdot \vec{\Psi})^2 
\ee
and
\be \label{case2}
\mL = i \vec{\Psi}^* {\cdot }\partial_t \vec{\Psi} - |\nabla \vec{\Psi}|^2 -\frac{\lam}{4N}  \vec{\Psi}^{\, 2} (\vec{\Psi}^*)^2~.
\ee
Starting with (\ref{case1}), 
at leading order in large $N$, the only effect of the interaction is that the sum of cactus diagrams generates a mass term $|\Psi|^2$. Since the propagator remains free, the quasi-classical approximation holds, and the kinetics are governed by the  quantum Boltzmann (kinetic) equation for the occupation numbers $n_{\bf p_1}$, 
\be
\(\frac{\partial }{\partial t} + \mathbf{v}_1 \cdot \nabla \)n_{\bf p_1} = I_{\text{coll}}[n]
\ee
where the collision integral is, see e.g. \cite{Kolb, Liboff},
\bml \label{Boltzmann}
I_{\text{coll}}[n] = N \int \frac{d^d p_2}{(2\pi)^d} \frac{d^d p_3}{(2\pi)^d} \frac{d^d p_4}{(2\pi)^d} \, \vert{}\mA\vert{}^2 (2\pi)^{d} \delta(\mathbf{p}_1 {+ }\mathbf{p}_2 {-} \mathbf{p}_3 {-} \mathbf{p}_4) \,2\pi\delta(\omega_{\bf p_1} {+} \omega_{\bf p_2} {-} \omega_{\bf p_3} {- } \omega_{\bf p_4}) \\
\Big[ n_{\bf p_3} n_{\bf p_4} (1 + n_{\bf p_1})(1 + n_{\bf p_2}) - n_{\bf p_1} n_{\bf p_2} (1 + n_{\bf p_3})(1 + n_{\bf p_4}) \Big]~,
\end{multline}
where  $\omega_{\mathbf{p}} = p^2$ for the nonlinear Schr\"{o}dinger case. Here $\mA$ is the two to two scattering amplitude with fixed large $N$ indices. In the kinetic equation for a fixed component, the index of its scattering partner is summed over, producing the overall factor of $N$.

\begin{figure}
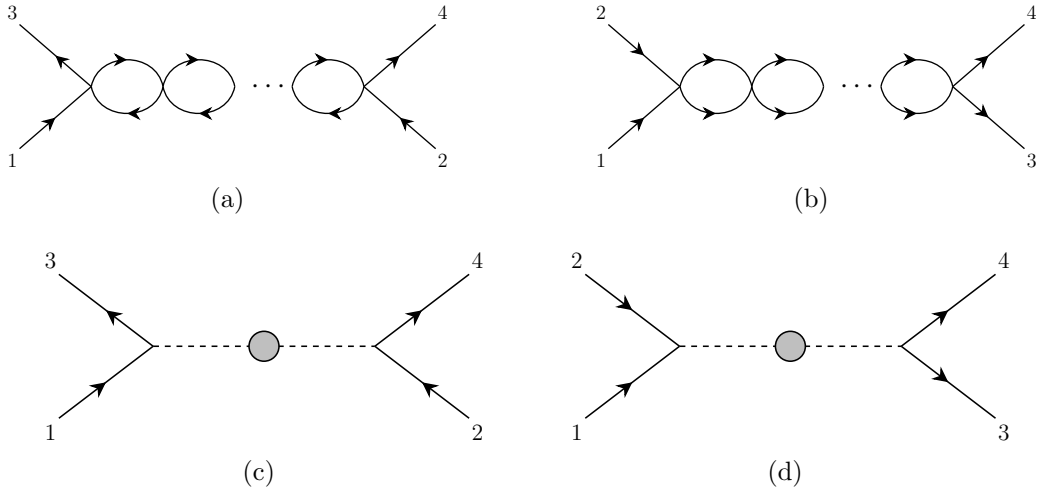

\centering
\subfloat[]{\BubbleChainOne{6cm}} \ \ \ \ \ \  \ \ \ \ \ \ 
\subfloat[]{
\BubbleChainTwo{6cm}} \\
\subfloat[]{
\HSSdiagram{6cm}} \ \ \ \ \ \ 
\subfloat[]{
\HSdiagram{6cm}} 
\caption{The large-$N$ four-point functions for (\ref{case1}, \ref{case2}) are given by the sum of bubble diagrams, (a), (b), respectively. This can be viewed as the exchange of a Hubbard-Stratonovich $\phi$ particle, (c), (d), respectively.  } \label{BubbleSum}
\end{figure}

The large $N$ connected four-point function, which scales as $1/N$, is given by a sum of bubble diagrams, see Fig.~\ref{BubbleSum}. However, since we are not in the vacuum, it is not immediately clear what $\mA$ is. A Hubbard-Stratonovich transformation clarifies this:  we introduce an auxiliary field $\phi$ with terms in the Lagrangian $\frac{N}{2\lambda}\phi^2-\phi |\v \Psi|^2 $, so integrating out $\phi$ gives back $-\frac{\lambda}{2N}(|\v \Psi|^2)^2 $.
The scattering amplitude is now clear: it is simply the retarded propagator $D^R$ of $\phi$.~\footnote{See Appendix~\ref{appA1} for a more systematic derivation from the full Dyson equations on the Keldysh contour.}  Accounting for the $i$ in the exponent of the path integral, the $\phi$ field has a bare propagator of $i\lambda/N$ times a delta function. The full (matrix) propagator $D$ of $\phi$ is given by a sum of bubble diagrams, see Figs.~\ref{BubbleSum}(a), (c), where a single bubble is given by the function $\Pi$, 
\be \label{25}
D(\o,\bf k) = \frac{i\lam/N}{1 + i\lam \Pi(\o,\bf k)} ~, \ \ \ \ \Pi(1,2) = G(1,2)G(2,1)~, \  \ \  \Pidiagram{3cm}
\ee
where the first equation is in momentum space and the second is in position space. The retarded component of $D$ in momentum space is
\be
D^R(\o,\bf k)= \frac{i\lam/N}{1 + i\lam \Pi^R(\o,\bf k)}\label{DR mom independent}~.
\ee
where $\Pi^R(\o,\bf k)$, written as a convolution, is
\be \label{27}
\Pi^R(\o, \mathbf{k}) = \frac{1}{2} \int \frac{d\nu d^d q}{(2\pi)^{d+1}} \left[ G^R(\nu, \mathbf{q}) G^K(\nu {- }\o, \mathbf{q} {-} \mathbf{k}) - G^K(\nu, \mathbf{q}) G^A(\nu{ -} \o, \mathbf{q} {-} \mathbf{k}) \right]~.
\ee
Evaluating with free propagators (\ref{A13}) gives, 
\be \label{PiR}
\Pi^R(\omega, \mathbf{k}) = i\int \frac{d^d q}{(2\pi)^d} \frac{n_{\mathbf{q}-\mathbf{k}} {-}n_{\mathbf{q}} }{\omega {-} \omega_{\mathbf{q}} {+} \omega_{\mathbf{q}{-}\mathbf{k}} {+} i\epsilon}~.
\ee
The kinetic equation is given by (\ref{Boltzmann}), where the amplitude is the propagator $D^R$ of a $t$-channel exchanged $\phi$ particle: it has energy $\o_1 - \o_3$ and momentum $\bf p_1-\bf p_3$, 
\be \label{ADR}
\mA = D^R(\o = p_1^2{-}p_3^2, \bf p_1{ -}\bf p_3)~.
\ee
This recovers the large $N$ kinetic equation  in the literature \cite{Kraichnan, Nowak:2011sk,B15,Gaz19,bergesGasenzerScheppach2010, berges2002,bergesRothkopfSchmidt2008, Rosenhaus:2024iqw, RF2, Rosenhaus:2025tjx}.
From the kinetic equation one can derive the turbulent solutions for an inverse cascade, both in the UV (weak coupling) and IR (strong coupling). One finds the turbulent solutions (for this large $N$ theory) approach critical balance scaling in the IR for negative  coupling case, realizing, for the first time, a strong turbulence scaling exponent that has long been conjectured on heuristic grounds \cite{Phil, Goldreich, Newell, NS}. Direct numerical simulations in 2d reveal that  the same critical balance scaling is approached in the IR for the $N=1$ nonlinear Schr\"odinger equation \cite{Rosenhaus:2025tjx}. 

Consider now instead the interaction (\ref{case2}). The kinetic equation is again given by (\ref{Boltzmann}), where the amplitude is the propagator $D^R$, but now of an $s$-channel exchanged complex $\phi$ field: it has energy $\o_1+\o_2$ and momentum $\bf p_1+\bf p_2$, see Figs.~\ref{BubbleSum} (b),  (d),
\be \label{ADR2}
\mA = D^R(\o = p_1^2+p_2^2, \bf p_1+\bf p_2)~,
\ee
where now $\Pi(1,2) = \frac{1}{2}G(1,2)^2$ and hence
\be \label{211}
\Pi^R(\omega, \mathbf{k}) = \frac{i}{2}\int \frac{d^d q}{(2\pi)^d} \frac{n_{\mathbf{q}} {+} n_{\mathbf{k}-\mathbf{q}}{+}1}{\omega { -}\omega_{\mathbf{q}} {-} \omega_{\mathbf{k}-\mathbf{q}} {+ }i\epsilon}~.
\ee
In the classical wave limit, $n_{\bf q} \gg 1$, the $1$ on the right-hand side can be dropped. 

\subsection{Momentum-dependent interaction}\label{Sec mom dep}
In this section we take Hamiltonians which are generalizations of (\ref{case1}) and (\ref{case2}), respectively, to a general dispersion relation $\o_p$ and, more importantly, to a general momentum-dependent quartic interaction:
\be
H = \int \! \frac{d^dp}{(2\pi)^d}\, \o_p |\v \Psi( p)|^2+ H_{\text{int}}~,
\ee
where the interaction is respectively either
\be \label{Hm2}
H_{\text{int}} = \frac{1}{2N}\int \prod_{i=1}^4 \frac{d^d p_i}{(2\pi)^d}\lam_{\bf p_1\bf  p_2 \bf p_3 \bf p_4} \v \Psi(p_1)^*{\cdot}\v \Psi (p_3) \v \Psi(p_2)^*{\cdot}\v \Psi(p_4)  (2\pi)^d\delta(\bf p_1{+} \bf p_2{-}\bf p_3 {-}\bf p_4)~,
\ee
or
\be \label{Hm1}
H_{\text{int}} = \frac{1}{4N}\int\prod_{i=1}^4 \frac{d^d p_i}{(2\pi)^d} \lam_{\bf p_1\bf  p_2 \bf p_3 \bf p_4}\!  \v \Psi(p_1){\cdot}\v \Psi (p_2) \v \Psi^*(p_3){\cdot}\v \Psi^*(p_4) (2\pi)^d\delta(\bf p_1{+} \bf p_2{-}\bf p_3 {-}\bf p_4)~.
\ee

We start with the $s$-channel interaction (\ref{Hm1}). Our task is to sum the bubble diagrams. Because of the momentum dependence, the bubble integrals do not decouple; we must instead write an integral equation for their sum. It is  
useful to work with the conserved momentum $\mathbf{Q}$ running through a bubble chain: we shift to $(\Omega, \bf Q)$ where $\Omega = \o_1 + \o_2$ and $\bf Q = \bf p_1 + \bf p_2 = \bf p_3 + \bf p_4$  and: 
\be \label{kk}
\text{ incoming } \mathbf{k} = (\mathbf{p_1} - \mathbf{p_2})/2~,\ \ \ \ \text{ outgoing } \mathbf{k'} = (\mathbf{p_3 }- \mathbf{p_4})/2~.
\ee
 We can re-parameterize the bare vertex as $\lambda_{\bf k\,\bf  k'}(\mathbf{Q}) \equiv \lambda_{\bf Q/2+\bf k, \bf Q/2-\bf k, \bf Q/2+\bf k', \bf Q/2-\bf k'}$. The dressed propagator is similarly parameterized as $D_{\bf k\, \bf k'}(\Omega, \mathbf{Q})$. Looking at the retarded component, we have the integral equation,~\footnote{ This result matches what appeared in \cite{FR}, see  arXiv v2, Eq. B59; it was not recognized at the time that  only  $D^R$, and nothing more, is needed to get the kinetic equation.} 
\be \label{IntEQ}
D^R_{\mathbf{k}\, \mathbf{k}'}(\Omega, \mathbf{Q}) = \frac{i}{N}\lambda_{\mathbf{k}\, \mathbf{k}'}(\mathbf{Q}) + \frac{1}{2}\int \frac{d^d p}{(2\pi)^d} \, \lambda_{\mathbf{k}\, \mathbf{p}}(\mathbf{Q}) \left(\frac{1 + n_{\mathbf{Q}/2 + \mathbf{p}} + n_{\mathbf{Q}/2 - \mathbf{p}}}{\Omega- \o_{\mathbf{Q}/2+\mathbf p} - \o_{\mathbf{Q}/2-\mathbf p}  + i\epsilon} \right) D^R_{\mathbf{p}\, \mathbf{k}'}(\Omega, \mathbf{Q})~.
\ee
For the dispersion relation $\o_p =p^2$, the denominator simplifies to $\Omega - \o_{\mathbf{\frac{Q}{2} +p}} - \o_{\mathbf{\frac{Q}{2} -p}}  
= 2(\mathbf{k}^2 - \mathbf{p}^2)$. 
The kinetic equation is (\ref{Boltzmann}) with the scattering amplitude given by the on-shell retarded propagator of $\phi$,  $\mA =D^R_{\mathbf{k} \mathbf{k}'}(\omega_{\mathbf{p}_1}+\omega_{\mathbf{p}_2}, \mathbf{Q})$.

Now we turn to finding the amplitude in the case of the $t$-channel interaction in (\ref{Hm2}). One distinction from (\ref{Hm1}) is that at leading order in $1/N$, the cactus diagrams will shift the dispersion relation: the free value $\o_{\bf k}$ is replaced by \cite{RS1}
\be \label{RenormalizedOmega}
\o_{\bf k} \rightarrow \o_{\bf k} + \int \frac{d^dq}{(2\pi)^d} \lam_{ \bf k\bf q\bf k\bf q}(2n_{\bf q} {+}1)~.
\ee
To write the integral equation for the $\phi$ propagator, we work with the conserved transferred energy-momentum $Q = (\Omega, \mathbf{Q})$, where $\Omega = \omega_1 - \omega_3$ and $\mathbf{Q} = \mathbf{p}_1-\mathbf{p}_3 = \mathbf{p}_4-\mathbf{p}_2$, 
and re-parameterize the interaction coefficient as
$
	\lambda_{\mathbf{p}\,\mathbf{p}'}(\mathbf{Q})= \lambda_{\mathbf{p}+\frac{\mathbf{Q}}{2},\, \mathbf{p}'-\frac{\mathbf{Q}}{2},\, \mathbf{p}-\frac{\mathbf{Q}}{2},\, \mathbf{p}'+\frac{\mathbf{Q}}{2}}
$
with $\mathbf{p}=\left(\mathbf{p}_1+\mathbf{p}_3\right)/2$ and $\mathbf{p}'=\left(\mathbf{p}_2+\mathbf{p}_4\right)/2$.  The transpose of $\lambda$ is given by $\lambda_{\mathbf{p}'\,\mathbf{p}}(\mathbf{Q}) = \lambda_{\mathbf{p}\,\mathbf{p}'}(-\mathbf{Q})$, and $\lambda$ at fixed $\mathbf{Q}$ is Hermitian, $\lambda_{\mathbf{p}'\,\mathbf{p}}(\mathbf{Q})^\star = \lambda_{\mathbf{p}\,\mathbf{p}'}(\mathbf{Q})$.
 The integral equation for the retarded dressed interaction $D^R_Q$ is \cite{Rosenhaus:2024iqw}
 \be \label{IntEQ2}
D^R_{\mathbf{k}\, \mathbf{k}'}(\Omega, \mathbf{Q}) = \frac{i}{N}\lambda_{\mathbf{k}\, \mathbf{k}'}(\mathbf{Q}) + \int \frac{d^d p}{(2\pi)^d} \, \lambda_{\mathbf{k}\, \mathbf{p}}(\mathbf{Q}) \left( \frac{n_{\mathbf{p}-\mathbf{Q}/2} - n_{\mathbf{p}+\mathbf{Q}/2}}{\Omega - \omega_{\mathbf{p}+\mathbf{Q}/2} + \omega_{\mathbf{p}-\mathbf{Q}/2} + i\epsilon} \right) D^R_{\mathbf{p}\, \mathbf{k}'}(\Omega, \mathbf{Q})~.
\ee
The kinetic equation is (\ref{Boltzmann}) with $\o_{\bf k}$ given by (\ref{RenormalizedOmega}) and with the scattering amplitude given by the on-shell retarded propagator of $\phi$,  $\mA =D^R_{\mathbf{k} \mathbf{k}'}( \omega_{\mathbf{p}_1}{-}\omega_{\mathbf{p}_3}, \mathbf{Q})$.

\subsection{Special class}
There is a class of interactions for which the integral equation determining the $\phi$ propagator can be explicitly solved. In particular, 
in the case of the $s$-channel interaction (\ref{Hm1}) it takes the form
\be 
 \lam_{\bf p_1 \bf p_2 \bf p_3\bf p_4} = \sum_{i,j=1}^M \Lambda_{ij} f_i(\bf p_1,\bf p_2) f_j(\bf p_3, \bf p_4)~,
\ee
where $\Lambda_{ij}$ is a constant Hermitian matrix and $f_i$ are arbitrary functions, 
while for the $t$-channel interactions in (\ref{Hm2})  it takes the form, 
\be
 \lam_{\bf p_1 \bf p_2 \bf p_3\bf p_4} = \sum_{i,j=1}^M \Lambda_{ij} f_i(\bf p_1,\bf p_3) f_j(\bf p_2,\bf p_4)~,
\ee
 We may consider both cases by using our re-parameterized interaction, 
\begin{align} \label{fac1}
 \lambda_{\mathbf{k}\,\mathbf{k}'}(\mathbf{Q}) = \sum_{i,j=1}^M f_i(\mathbf{k}; \mathbf{Q} )^*\Lambda_{ij} f_j(\mathbf{k}'; \mathbf{Q} )~.
\end{align}
We can express the dressed interaction by assuming it retains the same separable form as the bare interaction:
\begin{align}
	\label{Dm}
	D^R_{\mathbf{k}\, \mathbf{k}'}(\Omega, \mathbf{Q}) = \sum_{i,j=1}^M f_i(\mathbf{k}; \mathbf{Q} )^* D^R_{ij}(\Omega, \bf Q) f_j(\mathbf{k}'; \mathbf{Q} )~.
\end{align}
Inserting into either (\ref{IntEQ}) or (\ref{IntEQ2}) and factoring out the basis functions $f_i(\bf k)$ and $f_j(\bf k')$ on the outside yields a  matrix equation for the coefficients:
\be \label{DijRG}
D^R_{ij} = \frac{i}{N}\Lambda_{ij} -i \sum_{m,n=1}^M \Lambda_{im} \Pi_{mn}^R D_{nj}^R~,
\ee
where  in the $s$-channel interaction case (\ref{IntEQ}) 
\be
\Pi_{ij}^R(\Omega, \bf Q) = \frac{i}{2}\int \frac{d^d q}{(2\pi)^d} \, f_i(\mathbf{q}; \mathbf{Q}) \left(\frac{1 + n_{\mathbf{Q}/2 + \mathbf{q}} + n_{\mathbf{Q}/2 - \mathbf{q}}}{\Omega- \o_{\mathbf{Q}/2+\mathbf q} - \o_{\mathbf{Q}/2-\mathbf q}  + i\epsilon} \right) f_j(\mathbf{q}; \mathbf{Q})^*.
\ee
and in the $t$-channel interaction case (\ref{IntEQ2})
\be
\Pi_{ij}^R(\Omega, \bf Q) = i\int \frac{d^d q}{(2\pi)^d} \, f_i(\mathbf{q}; \mathbf{Q})  \left( \frac{n_{\mathbf{q}-\mathbf{Q}/2} - n_{\mathbf{q}+\mathbf{Q}/2}}{\Omega - \omega_{\mathbf{q}+\mathbf{Q}/2} + \omega_{\mathbf{q}-\mathbf{Q}/2} + i\epsilon} \right)f_j(\mathbf{q}; \mathbf{Q})^*,
\ee
In  matrix notation we may write (\ref{DijRG}) as 
\be
\mathbf{D}^R = \frac{i}{N}\boldsymbol{\Lambda} - i \boldsymbol{\Lambda} \mathbf{\Pi}^R \mathbf{D}^R~,
\ee
which we may solve to obtain $D^R(\Omega, \bf Q)$, 
\be \label{Dmatrix}
\mathbf{D}^R(\Omega, \bf Q) = \frac{i}{N}\left[ \mathbf{I} + i\boldsymbol{\Lambda} \mathbf{\Pi}^R(\Omega, \bf Q) \right]^{-1} \boldsymbol{\Lambda}~.
\ee
Note that in the momentum-independent case, where there is a single component and $f = 1$, this indeed reduces to the earlier expression \eqref{DR mom independent} with $\Pi^R$ given by \eqref{PiR} or \eqref{211}, respectively.

\subsection*{Examples}
Let us look at some examples. If we take $M=1$, we have an interaction of the product factorized form,
\be \label{lamProd}
  \lam_{\bf{p}_1 \bf{p}_2 \bf{p}_3 \bf{p}_4} =\lam f_1(\bf{p}_1, \bf{p}_2) f_1(\bf{p}_3,\bf{p}_4)~, \ \ \ \ \ \lambda_{\bf k \, \bf k'}(\bf Q) = \lam f_1(\bf k; \bf Q)^* f_1(\bf k'; \bf Q)~,
 \ee
 where we have written it in both original and re-parameterized form, 
and (\ref{Dmatrix}) gives
\be
D_{\bf k \bf k'}^R(\Omega, \bf Q) =\frac{i}{N} \frac{\lambda_{\bf k \, \bf k'}(\bf Q)}{1 + i\lambda \Pi_{11}^R(\Omega, \bf Q)}~.
\ee
A notable case of this form is the MMT model \cite{MMT, Miguel}.

Taking $M=2$ now, we simply invert the $2\times 2$ matrix in (\ref{Dmatrix}) to get
\begin{align}
\bf D^R = \frac{i}{N}\frac{\bf\Lambda + i \,\text{det}(\bf \Lambda) \,\text{adj}\,({\bf \Pi^R})}{1+i\,\text{Tr}(\bf \Lambda \bf \Pi^R)-\text{det}(\bf \Lambda \bf \Pi^R)},\quad \text{adj}\,({\bf \Pi^R})\equiv 	\left(\begin{array}{cc}
			\Pi_{22}^R& -\Pi_{12}^R\\
		-\Pi_{21}^R & \Pi_{11}^R
		\end{array}\right).
\end{align}
A special case involves $\Lambda_{11} = \Lambda_{22} = 0$, $\Lambda_{12} = \Lambda_{21} = \lam$ and real $f_1 = f$ and $f_2 = 1$,
\be \label{lamSum}
\lam_{\bf{p}_1 \bf{p}_2 \bf{p}_3 \bf{p}_4} =\lam \(f(\bf{p}_1,\bf{p}_2) + f(\bf{p}_3, \bf{p}_4)\)~, \ \ \ \ \  \lam_{\bf k \, \bf k'}(\bf Q) = \lam \left(f(\bf k; \bf Q)+ f(\bf k'; \bf Q)\right).
\ee
Inserting this to get the components and writing out the full momentum-dependent retarded interaction using $D^R = \sum_{i,j} f_i D_{ij}^R f_j$ gives:
\be
D^R_{\bf k\, \bf k'}(\bf Q) =\frac{i}{N} \frac{(1 +i\lambda \Pi_{12}^R)\lam_{\bf k \, \bf k'}(\bf Q)-i\lambda^2 \Pi_{22}^R f(\bf k; \bf Q)f(\bf k'; \bf Q) -i \lambda^2 \Pi_{11}^R}{(1 + i\lambda \Pi_{12}^R)^2 + \lambda^2 \Pi_{11}^R \Pi_{22}^R}~.
\ee
A special case of this type of interaction is for spin waves, which involves an $s$-channel interaction and for which $f(\bf{p}_1,\bf{p}_2) = -\bf{p}_1 {\cdot} \bf{p}_2$ (equivalently, $f(\bf{k}; Q) =- \bf{k}^2{-}\bf{Q}^2/4$). This was studied in \cite{Liu:2025rgt}, using the notation $\mL_0 = -4i\Pi_{22}^R,\,\mL_1=-4i \Pi_{12}^R,\,\mL_2= -4i \Pi_{11}^R,$ and taking the overall normalization $\lambda=4$.  

\subsection*{Non-resonant cubic interaction}

There is a context in which a factorized interaction of the form (\ref{fac1})  arises naturally: in non-resonant three-wave interactions. Namely, consider a cubic interaction between two fields, a complex vector $\v a_{\bf{k}}$ and a complex scalar $b_{\bf{k}}$, with Hamiltonian $H = H_0 + H_{int}$
\begin{gather}
	H_0 =\int \frac{d^d\bf{k}}{(2\pi)^d}\, \(\omega_{a}(\bf{k})|\v a_{\bf{k}}|^2 + \omega_{b}(\bf{k})|b_{\bf{k}}|^2 \),\\
	H_{int} =\frac{1}{\sqrt{4N}} \int\prod_{i=1}^3\frac{d^d \bf{p}_i}{(2\pi)^d} \left[ \lam_{\bf{p}_3; \bf{p}_1, \bf{p}_2} b_{\bf{p}_3}^\dagger\,  \v a_{\bf{p}_1}{\cdot} \v a_{\bf{p}_2} (2\pi)^d\delta(\bf{p}_3 {-} \bf{p}_1 {-} \bf{p}_2) + \text{c.c.} \right]~.
\end{gather}
If the resonance condition $\omega_b(\bf{p}_3) = \omega_a(\bf{p}_1) + \omega_a(\bf{p}_2)$ cannot be satisfied, then the leading order term in the kinetic equation vanishes, and we may look at the effective four-wave interaction of the $\v a$ fields; it takes the form (\ref{Hm1}) 
with
\begin{align}
	\lam_{\bf{p}_1 \bf{p}_2 \bf{p}_3 \bf{p}_4} = \frac{\lambda^*_{\bf Q; \bf p_3\bf p_4}\lambda_{\bf Q; \bf p_1\bf p_2}}{\Omega-\omega_b(\bf Q)}~,
\end{align}
The interaction appears within $D^R(\Omega, \bf Q)$ and it is evaluated on the four-wave resonance shell, so $\Omega = \omega_a(\bf p_1)+\omega_a(\bf p_2) = \omega_a(\bf p_3)+\omega_a(\bf p_4)$ and $\bf Q = \bf p_1 +\bf p_2 = \bf p_3 +\bf p_4$. Thus this is a factorizable $s$-channel interaction of the form (\ref{lamProd}).

Alternatively, we may take a cubic interaction of the form
\be
H_{int} = \frac{1}{\sqrt{2N}}\int \prod_{i=1}^3\frac{d^d \bf{p}_i}{(2\pi)^d} \left[ \lambda_{\bf{p}_3, \bf{p}_1, \bf{p}_2} b_{\bf{p}_3} \v a_{\bf{p}_1}^{\, \dagger} {\cdot} \v a_{\bf{p}_2} (2\pi)^d \delta(\bf{p}_1 - \bf{p}_2 - \bf{p}_3) + \text{c.c.} \right]~,
\ee
which, if the resonance condition $\o_b(\bf{p}_3) = \o_a(\bf{p}_1)-\o_a(\bf{p}_2)$ cannot be satisfied, gives an effective quartic interaction of the form (\ref{Hm2}),
\be
\lam_{\bf{p}_1 \bf{p}_2 \bf{p}_3 \bf{p}_4} =  \frac{ \lambda_{\bf{Q}, \bf{p}_4, \bf{p}_2}^*\lambda_{\bf{Q}, \bf{p}_1, \bf{p}_3}}{\Omega-\omega_b(\bf Q)}~,
\ee
which is a factorizable $t$-channel interaction of the form \eqref{lamProd}, with $\Omega = \omega_a(\bf p_1)-\omega_a(\bf p_3) = \omega_a(\bf p_4)-\omega_a(\bf p_2)$ and $\bf Q = \bf p_1 -\bf p_3 = \bf p_4 -\bf p_2$.  A prime example is in Langmuir-ion acoustic turbulence: a high frequency Langmuir (electron plasma) wave interacting with a low frequency ion-acoustic (sound) wave. This will be studied elsewhere.

\section{Random coupling model} \label{Sec3}
In this section we consider another large $N$ model. It has the same field content as the model in the previous section: $N$ complex scalars $\Psi_i(x)$ with a quartic interaction. The interaction has  Gaussian random couplings, $J_{ijkl}$, with zero mean and variance $\overline{ J_{ijkl}^2}= J^2/8N^3$ for each $i, j,k,l$ drawn from $1$ to $N$.~\footnote{One can naturally produce a random coupling from a Clebsch-Gordan coefficient \cite{Amit:1979ev, PhysRevLett.70.1101,PhysRevE.49.3990, Benedetti:2020iku, Biggs:2026zir}. Alternatively, one can study a tensor model  \cite{Gurau:2016lzk, Witten:2016iux,Klebanov:2016xxf}, which has the same large $N$ properties as the random coupling theory.} The Hamiltonian and corresponding Lagrangian density are, 
\be
H = \int d^d x \left( \vert{}\partial_x \Psi_i\vert{}^2 + J_{i j k l} \Psi_i^* \Psi_j^* \Psi_k \Psi_l \right)~, \ \ \ \ \  \ \mL = \Psi_i^* (i \partial_t + \nabla^2) \Psi_i - J_{i j k l} \Psi_i^* \Psi_j^* \Psi_k \Psi_l~.
\ee
Large $N$ random coupling models date back to Kraichnan  \cite{Hansen,KraichnanSYK, Betchov, Hu:2023aky}, who introduced and solved them (before either large $N$ or the Keldysh contour were invented). Their most well-known realization is in the form of the SYK model \cite{Kitaev,SY, Rosenhaus:2018dtp}. Since our field $\Psi$ is a scalar in higher dimensions, this model is most similar to the $d$-dimensional bosonic SYK model \cite{Liu:2018jhs}. Since $\Psi$ is a higher-dimensional scalar, this model closely resembles the $d$-dimensional bosonic SYK model \cite{Liu:2018jhs}. The complex field and single time derivative are minor differences; the analysis easily adapts to a real relativistic scalar."

The central feature of SYK is that at large $N$ the dominant diagrams are melons: the self-energy is 
\be \label{SigmaSYK}
\Sigma(1,2) = -J^2 G(1,2)^2 G(2,1)~, \ \ \ \ \ \ \ \ \SigmadiagramM{2cm}~.
\ee
Taking the Keldysh and retarded components and inserting into the kinetic equation (\ref{A10}), see Appendix~\ref{appA2}, gives, in the classical limit, 
\begin{multline} \label{SYKKE}
 \left( \partial_T + 2\mathbf{p}_1 \cdot \nabla_\mathbf{X}  \right)n_{\bf p_1}(T) =
  \frac{J^2}{4} \text{Re} \int_{0}^{T}\!\! dt' \!\! \int \frac{d^d p_2}{(2\pi)^d} \frac{d^d p_3}{(2\pi)^d} \frac{d^d p_4}{(2\pi)^d} (2\pi)^d \delta^{(d)}(\mathbf{p}_1 {+ }\mathbf{p}_2 {-} \mathbf{p}_3 {-} \mathbf{p}_4)  \\
   \Big[ \left(\rho^*_1 {G^K_2}^* + {G^K_1}^* \rho^*_2\right)G^K_3 G^K_4 -{G^K_1}^* {G^K_2}^*\left(\rho_3 G^K_4 + G^K_3 \rho_4\right) \Big]~.
\end{multline}
Here  $\rho$ is the spectral density \eqref{Def rho}, and $G^K$ is the Keldysh Green's function. We use the compact notation $	G^K_i \equiv G^K(\mathbf{X}, T, t'; \mathbf{p}_i)$, and similarly for $\rho$.

As a consistency check, if one substitutes the free propagator and spectral density on the right, the standard weak coupling wave kinetic equation is recovered. Each propagator $G^K_i$ provides a factor $n_i$, so we obtain the $(n_1+n_2)n_3 n_4 - n_1 n_2 (n_3+n_4)$ structure of the classical limit of \eqref{Boltzmann}. Both $G^K_i$ and $\rho_i$ also involve an exponential factor $\exp(-i\omega_{\mathbf{p}_i}(T-t'))$. After integration over $t'$, these will lead to a delta function enforcing the resonance condition.

Of course, here the propagators on the right-hand side are not the free propagators. As a result, this kinetic  equation is qualitatively different from the standard  kinetic equation, unlike what we found  for the large $N$ vector model. Namely, it  is not a closed equation for $n_k(T)$: the occupation numbers at time $T$ depend on the full propagators at  earlier times. The propagators must be determined self-consistently from the full Schwinger-Dyson equations (\ref{14}) with self-energy (\ref{SigmaSYK}), as in two-particle irreducible treatments of nonequilibrium field theory \cite{berges2002}. The spectral density \eqref{Eq rho gamma} exhibits an interaction-induced frequency shift and spectral broadening, and we may even consider a strong regime in which the self-energy $\Sigma^R$ dominates the bare dispersion relation.

\subsection{Scale invariance in the IR: strong turbulence}
We now look for stationary, scale-invariant solutions in the strong-coupling regime. For a stationary state, the propagators only depend on the relative time $\tau = t_1-t_2$. Let us assume a scale invariant ansatz, 
\be
G^K(\mu^{-z} \tau, \mu k) = \mu^{-\g} G^K(\tau, k)~,
\ee
where $z$ and $\g$ are scaling exponents to be determined. 
Notice that for equal times, $\tau=0$, $G^K(\tau=0,  k) = 2n_{\bf k}$, and so this reduces to $n_{\bf k} \sim k^{-\g}$. 
The retarded Green's function is a commutator, see (\ref{13}), and 
so is a constant when evaluated at equal times, $G^R(t_1=t_2=t, k)=1/2$. This means that $G^R$ behaves as
\be
G^R(\mu^{-z} \tau, \mu k) = G^R(\tau,k)~.
\ee
This implies $
G^K(\tau, k) = k^{- \g} \Phi^K(k^z\tau )$ and $G^R(\tau, k) = \Phi^R(k^z \tau )
$ 
for scaling functions $\Phi^K, \Phi^R$.

From the expression for   $\Sigma^R$ for melon diagrams (\ref{A19}) we get
\be \label{sigmaRre}
\Sigma^R(\mu^{-z} \tau, \mu k) = \mu^{2d-2\g} \Sigma^R(\tau,k)~, \ \ \ \ \ \ \Sigma^R(\mu^{z} \o, \mu k) = \mu^{2d-2\g-z} \Sigma^R(\o,k)~,
\ee
where in the second equality we translated into frequency space. Likewise, for the Green's functions in frequency space,
\be \label{GkGr}
G^K(\mu^z \o, \mu k) = \mu^{-z- \g} G^K(\o, k)~, \ \ \ G^R(\mu^{z} \o, \mu k) = \mu^{-z}G^R(\o,k)~.
\ee

Now comes the essential part: in the retarded propagator $G^R$ in (\ref{Eq GR}) we consider the infrared and strong coupling limit in which the term $\omega_\bf k$ may be neglected compared to $\Sigma^R$, 
\be
G^R(\mu^z \o, \mu k) = \frac{i}{\mu^z \o - i\Sigma^R(\mu^z \o, \mu k)}~.
\ee
Inserting (\ref{GkGr}) and (\ref{sigmaRre}) and canceling a $\mu^{-z}$ on both sides gives
\be
G^R(\o, k) = \frac{i}{\o - i\mu^{2d - 2\g -2 z} \Sigma^R(\o,k)}~.
\ee
From this we conclude that we must have
\be \label{zR}
2d - 2\g -2 z =0~ \ \ \ \Rightarrow z = d- \g~.
\ee

We derived this condition on $z$ by enforcing consistency between the scale-invariant ansatz and the strong-coupling self-energy. A second condition arises by requiring the ansatz to be a stationary solution with constant number flux $Q$.
The argument is a variation on the standard argument determining $\gamma$ for the inverse cascade in weak wave turbulence \cite{ZLF}.

The number flux $Q(p)$ is given by integrating the collision integral $I_{coll}(\bf p_1)$ on the right hand side of \eqref{SYKKE} up to some absolute value $p$,
\be
	Q(p) = -\int^p \frac{d^d p_1}{(2\pi)^d}I_{coll}(\bf p_1) \sim J^2 \int_{|\bf p_1|\leq p}  d\tau \left( d^d p_i\right)^4 \delta^{(d)}\left(\sum \bf p\right)\left(G^K\right)^3 \rho~.
\ee
The last term is in schematic form, making clear the scaling in the collision integral. By rewriting the integral in terms of the tilded variables $p_i = p \,\tilde{p}_i, \tau = p^{-z}\,\tilde\tau$, and using the scaling function form of $G^K$ and $ \rho$, it is clear that $
	Q(p)\propto J^2 p^{-z+3d-3\gamma} $.
So a stationary solution with constant number flux must have
\begin{align}
	\gamma = d-\frac{z}{3}~. \label{Eq gamma}
\end{align}
Furthermore we must have $G^K(\tau=0)\sim |Q|^{1/3}J^{-2/3}k^{-d+z/3}$. This is very similar to the ordinary inverse cascade Kolmogorov-Zakharov solution, but at weak-coupling the exponent $\alpha$ in the dispersion relation $\omega_\bf k \sim k^\alpha$ appears instead of $z$. This can be understood from the fact that at leading order one uses the free propagators $G_0^K(\tau)$ and $\rho_0(\tau)$ in the collision integral, and these involve $\tau$ dependence only in the scaling combination $\omega_\bf k \tau$. Indeed, our random coupling model, at leading order in $1/N$, simply renormalizes the propagator; the interaction is unchanged. 
Combining $\gamma = d - z/3$ with (\ref{zR}), $\gamma = d - z$ (\ref{zR}), gives
\be
\g =d~, \ \ \ z=0~.
\ee

To summarize, we seem to have an inverse turbulent cascade for a theory with a quartic interaction in both the weak- and strong-coupling regimes. For weak coupling in the UV where $\Sigma^R$ may be neglected compared to $\omega_\bf k$, there is standard weak wave turbulence with scaling  $n_k \sim (Q/J^2)^{1/3} k^{-d+2/3}$. At strong coupling in the IR, where $\Sigma^R \gg \omega_\bf k$, the consistency arguments given here suggest that the occupation number crosses over to $n_k \sim (Q/J^2)^{1/3} k^{-d}$.~\footnote{In $3d$, the scaling $k^{-d}$ may  need to be modified by a $\log$ factor, as a result of a $\log$ divergence in the collision integral, as occurs in a direct cascade in weak turbulence \cite{PhysRevLett.130.133001}.} Interestingly, a heuristic argument was previously presented for a strong turbulence  scaling of $d$  \cite{Grisha, RF2}, although the dependence on flux and coupling given in that argument  are different. We should caution that, as usual in wave turbulence theory, to establish that one has a true stationary solution of the kinetic equation, convergence of the collision integral has to be checked. The same holds here; indeed, in our case this is a more challenging task, as the collision integral involves the full propagator, whose form we do not know.

\section{Discussion}
This paper has two key results. 

First: for large $N$ vector models, the standard kinetic equation --- valid in the quasi-classical limit, for a dilute gas of particles or weakly interacting waves --- continues to hold, with the scattering amplitude given by the retarded propagator of the auxiliary Hubbard-Stratonovich field, which we computed for general, momentum-dependent, interactions. The novelty lies in accommodating momentum-dependent interactions, which are ubiquitous in physical systems. This result makes possible the study of  wave turbulence of   a number of physical systems (at least, their large-$N$ extensions), following the cascade from weak to strong interactions. 

Second: in general, away from weak coupling, the equation governing the occupation numbers of modes at time $t$ is not closed, and depends on the full propagator at all earlier times. This is true for a $d$-dimensional, large $N$, random coupling model. We were, nevertheless, able to find the putative scaling exponents for stationary and scale invariant solutions in the IR. 
The validity and applicability of these novel exponents to cascades in physical (small $N$) systems will  be studied elsewhere.

\section*{Acknowledgments} 
We thank G.~Falkovich, M.~Smolkin, A.~Kamenev for helpful discussions.
This work  is supported in part by NSF grants PHY-2209116 and PHY-2608451 and by BSF grant  2022113.

\appendix{}

\section{Kinetic equation from the Dyson equation on the Keldysh contour} \label{appA}

In this appendix we review how one can obtain the general kinetic equation (\ref{A10}), and then apply this to the vector model in Sec.~\ref{appA1} and random coupling model in Sec.~\ref{appA2}. 

\subsection*{Definitions}
The theories we consider throughout the paper  have a bare retarded propagator
\begin{align}
	G_0^R(\omega,\mathbf{k})
	=
	\frac{i}{\omega-\omega_{\mathbf{k}}+i\epsilon}~,
\end{align}
with $\o_{\mathbf{p}}$ indicating the dispersion relation. The Schwinger-Dyson equation then gives
\begin{align}
	G^R(\omega,\mathbf{k})=\frac{i}{\omega-\omega_{\mathbf{k}}-i\Sigma^R(\omega,\mathbf{k})}.\label{Eq GR}
\end{align}
The real and imaginary parts of $\Sigma^R$ have an interpretation as the spectral width $\Gamma$ and the frequency shift $\delta\omega$,
\begin{align}
	\Gamma(\omega,\mathbf{k})
	\equiv
	-{\rm Re}\,\Sigma^R(\omega,\mathbf{k})~,
	\qquad
	\delta\omega(\omega,\mathbf{k})
	\equiv
	-{\rm Im}\,\Sigma^R(\omega,\mathbf{k})~.\label{Def gamma domega}
\end{align}
To clarify this interpretation, it is useful to define the spectral density
\begin{align}
	\rho(\omega,\mathbf{k})
	&\equiv G^R(\omega,\mathbf{k})+G^A(\omega,\mathbf{k})=2\,{\rm Re}\,G^R(\omega,\mathbf{k})\label{Def rho}\\
	&= \frac{2\Gamma(\omega,\mathbf{k})}{(\omega-\omega_{\mathbf{k}}-\delta\omega(\omega,\mathbf{k}))^2+\Gamma(\omega,\mathbf{k})^2}\label{Eq rho gamma}~.
\end{align}
In the bare $\rho_0$, for which $\Gamma, \delta\omega \rightarrow 0$, this is a delta function $\rho_0 = 2\pi \delta(\omega-\omega_{\bf k})$.  Interactions induce a frequency shift $\omega_{\bf k} \rightarrow \omega_{\bf k}+\delta\omega$ and a broadening of the delta function according to $\Gamma$.

\subsection*{Dyson equation}
We take the Dyson equation (\ref{11}) and rewrite it as either $\(G_0^{-1} - \Sigma\) G = 1$ or $G\(G_0^{-1} - \Sigma\)  = 1$. Switching to position and time, and letting e.g. $1$ denote $(t_1, x_1)$, we write this as:
\bea
\int d3 \, {G_0^R}^{-1}(1, 3) G^K(3, 2) &=& \int d3 \left[ \Sigma^R(1, 3) G^K(3, 2) + \Sigma^K(1, 3) G^A(3, 2) \right]\\
\int d3 \, G^K(1, 3) {G_0^A}^{-1}(3, 2) &=& \int d3 \left[ G^R(1, 3) \Sigma^K(3, 2) + G^K(1, 3) \Sigma^A(3, 2) \right]~.
\eea
Assuming spatial homogeneity, we Fourier transform the spatial coordinates and work at fixed momentum $\mathbf{k}$. In our convention where $G_0^R(1,2) = \theta(t_1-t_2)e^{-i\omega_{\bf k}(t_1-t_2)}$, the bare inverse propagators are ${G_0^R}^{-1}(1,2) = (\partial_{t_1} + i\omega_k)\delta(t_1-t_2)$ and ${G_0^A}^{-1}(1,2) = {G_0^R}^{-1}(2,1)^*$.
	
Adding the two equations above and dividing by two gives:
\be \label{A3}
\frac{1}{2}(\partial_{t_1} + \partial_{t_2})G^K(1,2) = I_{\text{coll}}(1,2)~,
\ee
where the collision term $I_{\text{coll}}(1, 2)$ is:
\be
I_{coll}=\frac{1}{2}\int d3 \Big[ \Sigma^R(1, 3) G^K(3, 2) + \Sigma^K(1, 3) G^A(3, 2) + G^R(1, 3) \Sigma^K(3, 2) + G^K(1, 3) \Sigma^A(3, 2) \Big]~.
\ee
Setting $t_1 = t_2 =T$ and using $G^K(T, T; \mathbf{k}) = 2 n_{\mathbf{k}}(T)+1$, the left-hand side of \eqref{A3} is simply $ \partial_T n_\mathbf{k}(T)$.

Our conventions for the spectral density and spectral width are,
\bea \nn
\rho(t, t'; \mathbf{k}) &=& G^R(t, t'; \mathbf{k}) + G^A(t, t'; \mathbf{k})\\ 
\Gamma(t, t'; \mathbf{k}) &=& -\frac{1}{2} \left[ \Sigma^R(t, t'; \mathbf{k}) + \Sigma^A(t, t'; \mathbf{k}) \right].   \label{A6}
\eea
Taking advantage of the fact that $G^R(t', T)=G^A(T, t')=0$ for $t'<T$, and similarly for $\Sigma^{R}, \Sigma^A$, the time-domain collision integral may be written in terms of $\rho, \Gamma$,
\be \label{KEM}
\frac{\partial n_\mathbf{k}(T)}{\partial T} = \text{Re} \int_{0}^{T} dt' \left[ \Sigma^K(T, t'; \mathbf{k}) \rho^*(T, t'; \mathbf{k}) - 2\Gamma(T, t'; \mathbf{k}) G^{K*}(T, t'; \mathbf{k}) \right]~.
\ee
In the Markovian limit, where center-time dependence may be neglected over the memory integral, we extend the upper limit to infinity and Fourier transform with respect to the relative time. The $\text{Re}$ operator drops the principal value of the integral, reducing the convolution to a local product:
\be \label{collision integral markovian}
I_{\text{coll}}[n_{\mathbf{k}_1}] = \frac{1}{2} \int \frac{d\omega_1}{2\pi} \Big[\Sigma^K(\omega_1, \mathbf{k}_1) \rho(\omega_1, \mathbf{k}_1) - 2\Gamma(\omega_1, \mathbf{k}_1) G^K(\omega_1, \mathbf{k}_1) \Big]~.
\ee

If we drop the assumption of spatial homogeneity then in general the left-hand side of (\ref{A3}) cannot be simplified. If one assumes that $n_k(X)$ depends weakly on the center of mass position $X$ (specifically, $\vert{}\nabla_\mathbf{X} n_\mathbf{k}\vert{} \ll \vert{}\mathbf{k}\vert{} n_\mathbf{k}$), then one can write, 
\bea  \label{A10}
\( \partial_T+ \mathbf{v}_g \cdot \nabla_\mathbf{X} \)n_\mathbf{k}(\mathbf{X}, T)\!\!\! &{=}&\!\!\! I_{\text{coll}}(\mathbf{X}, T; \mathbf{k})  \\
 \!\!\! I_{\text{coll}}(\mathbf{X}, T; \mathbf{k})\!\!\! &{ =}& \!\!\! \text{Re} \int_{0}^{T}\!\!\! dt' \left[ \Sigma^K(\mathbf{X}, T, t'; \mathbf{k}) \rho^*(\mathbf{X}, T, t'; \mathbf{k}) -2\Gamma(\mathbf{X}, T, t'; \mathbf{k}) G^{K*}(\mathbf{X}, T, t'; \mathbf{k}) \right]~. \nn
\eea
where for the free propagator we took a general dispersion relation $\o(k)$ leading to the group velocity $v_g = \frac{d \o_k}{dk}$.

\subsection{Vector model} \label{appA1}
We now show how to produce the kinetic equation for the vector $O(N)$ model (\ref{case1}). We need to compute the terms appearing in (\ref{collision integral markovian}). 
As shown above, a Hubbard-Stratonovich transformation introduces the auxiliary field $\phi$ with propagator $D$, while the $\Psi$ propagator is $G$. The self-energy $\Sigma$ of our field $\Psi$ is,
\be
\Sigma(\omega_1, \mathbf{k}_1) =-\int \frac{d\omega_3 d^d k_3}{(2\pi)^{d+1}} D(\omega_1-\omega_3, \mathbf{k}_1-\mathbf{k}_3) G(\omega_3, \mathbf{k}_3)~, \ \ \ \ \ \Sigmadiagram{3cm} \label{selfenergy t channel}
\ee
Taking the Keldysh and retarded components gives, 
$$\Sigma^K(\omega_1, \mathbf{k}_1) = \frac{1}{2}\int \frac{d\omega_3 d^d k_3}{(2\pi)^{d+1}} \Big[ D^K(\omega_1-\omega_3, \mathbf{k}_1-\mathbf{k}_3) G^K(\omega_3, \mathbf{k}_3) +\rho_D(\omega_1-\omega_3, \mathbf{k}_1-\mathbf{k}_3) \rho(\omega_3, \mathbf{k}_3) \Big]$$
$$\Gamma(\omega_1, \mathbf{k}_1) =\frac{1}{4}\int \frac{d\omega_3 d^d k_3}{(2\pi)^{d+1}} \Big[ \rho_D(\omega_1-\omega_3, \mathbf{k}_1-\mathbf{k}_3) G^K(\omega_3, \mathbf{k}_3) + D^K(\omega_1-\omega_3, \mathbf{k}_1-\mathbf{k}_3) \rho(\omega_3, \mathbf{k}_3) \Big]~.$$
where $\rho$ is the spectral density (\ref{A6}) for $\Psi$ and $\rho_D = D^R+D^A$ is the spectral density for $\phi$. 

To evaluate the collision integral to $\order{1/N}$ it is sufficient to use the leading-order propagators, with any $\order{1}$ frequency renormalization absorbed into $\omega_k$,
\be \label{A13}
\rho(\omega, \mathbf{k}) = 2\pi \delta(\omega - \omega_{\bf k})~, \ \ \ \ G^R(\omega, \mathbf{k}) = \frac{i}{\omega - \omega_{\mathbf{k}} + i\epsilon}~, \ \ G^K(\omega, \mathbf{k}) = \rho(\omega, \mathbf{k}) (1 + 2n_{\mathbf{k}})~.
\ee
After substituting $\Sigma^K$ and $\Gamma$ in the collision integral \eqref{collision integral markovian}, and integrating over the frequencies $\omega_1, \omega_3$, we get the expression
\begin{align}
	I_{coll}(\mathbf{k}_1)=\frac{1}{2}\int \frac{d^d k_3}{(2\pi)^d}\left[D^K(\Omega, \mathbf{Q})\left(n_3 - n_1\right)-\rho_{D}(\Omega, \mathbf{Q})\left(2n_1 n_3+n_1+n_3\right)\right],\label{collision integral intermediate step}
\end{align}
where $\mathbf{Q}= \mathbf{k}_1-\mathbf{k}_3$, and $\Omega$ is the difference of the \emph{on-shell} frequencies $\omega_{\mathbf{k}_1}-\omega_{\mathbf{k}_3}$. It remains to calculate $D^K$ and $\rho_D$.

From the identity (\ref{14}) applied to the $\phi$ field, whose self-energy is $\Sigma_\phi=-N\Pi$, we have, 
\be
\label{DK}
D^K = D^R \Sigma^K_\phi D^A =N|D^R|^2 \Pi^K~.
\ee
Note that in these Schwinger-Keldysh conventions, there is a sign difference in the Keldysh component depending on whether a function on the Keldysh contour is decomposed as a propagator or as a self-energy (inverse propagator). Here we understand $\Sigma_\phi^K$ as a self-energy component, but $\Pi^K$ as a propagator component. $\Pi^K$ can be calculated from the Keldysh component of (the Fourier transform of) $\Pi(1,2) = G(1,2)G(2,1)$,
\begin{align}
\Pi^K(\omega, \mathbf{k}) &= \frac{1}{2} \int \frac{d\omega_q d^d q}{(2\pi)^{d+1}} \Big[ G^K(\omega_q, \mathbf{q}) G^K(\omega_q-\omega, \mathbf{q}-\mathbf k) -\rho(\omega_q, \mathbf{q}) \rho(\omega_q-\omega, \mathbf{q}-\mathbf k) \Big]\nonumber\\
&= \label{PiK} \int\frac{d^dk_2}{(2\pi)^d}\frac{d^dk_4}{(2\pi)^d}(2\pi)^d\delta^{(d)}\left(\mathbf{k}+\mathbf{k}_2-\mathbf{k}_4\right)2\pi \delta(\omega+\omega_{\mathbf{k}_2}-\omega_{\bf{k}_4})\,\left(2 n_{2}n_{{4}} + n_{2} + n_{{4}} \right)~.
\end{align}

For the spectral density $\rho_D$ we similarly apply the retarded and advanced Schwinger-Dyson equations
\be
\label{rhoD}
\rho_D = -D^R\left(\Sigma_\phi^R+\Sigma_\phi^A\right)D^A = N|D^R|^2 \rho_\Pi~.
\ee
Here $\rho_\Pi(\o, k)\equiv 2 \text{Re }\Pi^R(\o, k)$. 
Using the expression (\ref{27}) for $\Pi^R$, $\rho_\Pi$ is calculated to be
\begin{align}
	\rho_\Pi(\omega, \mathbf{k}) = 
\int\frac{d^dk_2}{(2\pi)^d}\frac{d^dk_4}{(2\pi)^d}(2\pi)^d\delta^{(d)}\left(\mathbf{k}+\mathbf{k}_2-\mathbf{k}_4\right)2\pi \delta(\omega{+}\omega_{\mathbf{k}_2}{-}\omega_{\mathbf{k}_4})\,\left(n_{2} - n_{{4}} \right)\label{rhoPi}~.
\end{align}
Inserting (\eqref{DK}, \eqref{PiK}, \eqref{rhoD}, \eqref{rhoPi}) into (\ref{collision integral intermediate step}), we recover the Boltzmann equation (\ref{Boltzmann}) with amplitude (\ref{ADR}). 

Finally, let us briefly comment on how this argument is extended for momentum-dependent coupling. The notation in terms of momentum indices introduced in Sec \ref{Sec mom dep} was chosen so that formulas generalize straightforwardly. In the self-energy \eqref{selfenergy t channel}, $D$ is replaced by the diagonal element $D_{\bf{k} \bf{k}}$ where $\bf k = (\bf k_1 + \bf k_3)/2$. Following the same steps as before, the formulas \eqref{DK} and \eqref{rhoD} are respectively generalized to
\begin{align}
	D^K_{\bf{k} \bf{k}} = N \int \frac{d^d q}{(2\pi)^d} |D^R_{\bf k \bf q}|^2 \Pi^K_\bf{q}~,\qquad \rho_{D,\,\bf{k} \bf{k}}=N \int \frac{d^d q}{(2\pi)^d} |D^R_{\bf k \bf q}|^2 \rho_\bf{\Pi, \,q}~,
\end{align}
where $\Pi^K_\bf{q}$ and $\rho_\bf{\Pi, \,q}$ refer to the integrands of \eqref{PiK} and \eqref{rhoPi}, respectively. Explicitly,
\begin{gather*}
	\Pi^K_\bf{q}(\Omega, \bf Q) = 2\pi \delta(\Omega+\omega_{\mathbf{k}_2}-\omega_{\bf{k}_4})\,\left(2 n_{2}n_{{4}} + n_{2} + n_{{4}} \right)~,\\
		\rho_\bf{\Pi, \,q}(\Omega, \bf Q) = 2\pi \delta(\Omega+\omega_{\mathbf{k}_2}-\omega_{\bf{k}_4})\,\left(n_{2} - n_{{4}} \right)~,
\end{gather*}
with $\bf k_2 = \bf q - \bf Q/2$ and $\bf k_4 = \bf q + \bf Q/2$. Following the same steps as before, we recover the Boltzmann equation \eqref{Boltzmann}, now with $\mathcal{A} = D^R_{\bf k \bf q}$.

As a technical comment, we note that the Hubbard-Stratonovich field $\phi_\mathbf{p}$ associated to the interaction \eqref{Hm2} is real in the sense that $\phi_\mathbf{p}(\mathbf{Q})^* = \phi_\mathbf{p}(-\mathbf{Q})$. It appears in the Lagrangian as
\begin{align}
 \int \frac{d^dQ}{(2\pi)^d} \left[\frac{1}{2}\int \frac{d^dp \,d^dp'}{(2\pi)^{2d}} \,\phi_\mathbf{p}^* \left(N\lambda_{\mathbf{p}\,\mathbf{p}'}^{-1}\right) \phi_{\mathbf{p}'} - \int \frac{d^dp}{(2\pi)^{d}} \left(\Psi^\star _{\mathbf{p}+\frac{\mathbf{Q}}{2}}\cdot \Psi_{\mathbf{p}-\frac{\mathbf{Q}}{2}}\right)\phi_\mathbf{p}\right]~,
\end{align}
where the formal inverse $\lambda^{-1}$ is defined by  $\int \frac{d^d q}{(2\pi)^d}\, \lambda_{\mathbf{p\, q}}(\mathbf{Q}) \lambda^{-1}_{\mathbf{q\, p}'}(\mathbf{Q}) = (2\pi)^d \delta(\mathbf{p}-\mathbf{p}')$.

\subsection{Random coupling model} \label{appA2}
Here we fill in the details of the kinetic equation for the random coupling model. 
Taking the Keldysh and retarded components of the self-energy (\ref{SigmaSYK}) gives
\bea 
\Sigma^K(1, 2) &=& \frac{J^2}{4} \Big[ G^K(1, 2)^2 G^K(2, 1) - 2 G^K(1, 2) \rho(1, 2) \rho(2, 1) + \rho(1, 2)^2 G^K(2, 1) \Big]\\
\Sigma^R(1, 2) &=& -\frac{J^2}{4} \Big[ 2 G^K(1, 2) G^K(2, 1) G^R(1, 2) - G^K(1, 2)^2 G^A(2, 1) - G^R(1, 2) \rho(1, 2) G^A(2, 1) \Big]~. \nn
\eea
In the classical limit in which for the action on the Keldysh contour one only keeps the leading order term for the ``quantum'' field which is the difference of the field on the upper and lower branches (see e.g. \cite{Hu:2025bqi}). This amounts to retaining the terms with the maximal number of $G^K$ factors,
\bea \label{SigmaK}
\Sigma^K(1,2) &=&\frac{J^2}{4} G^K(1,2)^2 G^K(2,1)~, \ \ \ \ \ \ \ \ \ \ \ \SigmadiagramMM{2.5cm} \\ \label{SigmaR}
\Sigma^R(1,2) &=& -\frac{J^2}{4} \( 2G^K(1,2)G^K(2,1)G^R(1,2) - G^K(1,2)^2 G^A(2,1) \)~,\\ \nn
 &&\ \  \  \ \ \ \SigmadiagramMMM{2.5cm} \ \ \ \ \ \  \ \ \   \ \ \ \ \ \  \ \ \ \ \ \ \  \SigmadiagramMMMM{2.5cm}
\eea
Fourier transforming the relative position in the expression for the retarded self-energy, and denoting the relative time $\tau= T-t'$
\bml \label{A19}
\Sigma^R(\tau, \mathbf{k}) = -\frac{J^2}{4} \int \frac{d^d p}{(2\pi)^d} \frac{d^d q}{(2\pi)^d} \frac{d^d l}{(2\pi)^d} (2\pi)^d \delta^{(d)}(\mathbf{k}+\mathbf{l}-\mathbf{p}-\mathbf{q})\\ \left[ 2 G^K(\tau, \mathbf{p}) G^K(-\tau, \mathbf{l}) G^R(\tau, \mathbf{q}) - G^K(\tau, \mathbf{p}) G^K(\tau, \mathbf{q}) G^A(-\tau, \mathbf{l}) \right].
\end{multline}
This expression makes clear the scaling of $\Sigma^K$ \eqref{sigmaRre} in the scale invariant ansatz.

The general kinetic equation (\ref{A10}) involves the spectral width $\Gamma = - \text{Re}\, \Sigma^R$. Using the classical limit of $\Sigma^R$,
\begin{align}
	\Gamma(1,2)= \frac{J^2}{8}\left(2 G^K(1, 2) G^K(2, 1) \rho(1, 2) - G^K(1, 2)^2 \rho(2, 1)\right).
\end{align}
Substituting $\Gamma$ and $\Sigma^K$ in the collision integral \eqref{KEM},
\begin{multline}
	I_{coll}=\frac{J^2}{4}\text{Re} \int_{0}^{T} dt' \left[\rho(1,2)^*G^K(2,1)G^K(1,2)^2 \right.\\
	\left.- G^K(1,2)^*\left(2 G^K(2, 1) G^K(1, 2) \rho(1, 2) - \rho(2, 1)G^K(1, 2)^2 \right) \right].
\end{multline}
Here the time components $t_1=T$ and $t_2=t'$. Upon Fourier transforming the relative position, we recover the form of the collision integral \eqref{SYKKE} used in the main body. Of course, at leading order in $J$, we are allowed to replace the propagator and spectral density by their free values (\ref{A13}), and if we do this in the Markovian form of the collision integral \eqref{collision integral markovian}
we recover the standard weak wave kinetic equation.

\bibliographystyle{utphys}

\end{document}